\PassOptionsToPackage{dvipdfmx}{graphicx}
\documentclass[%
 aip,
 amsmath,amssymb,
 reprint,%
onecolumn
]{revtex4-1}

\usepackage{graphicx}%
\usepackage{dcolumn}%
\usepackage{bm}%

\usepackage[utf8]{inputenc}
\usepackage[T1]{fontenc}
\usepackage{mathptmx}
\usepackage{etoolbox}
\usepackage{amssymb}
\usepackage{lipsum}

\usepackage{caption}
\usepackage{caption}
\usepackage{here}
\makeatletter
\def\@email#1#2{%
 \endgroup
 \patchcmd{\titleblock@produce}
  {\frontmatter@RRAPformat}
  {\frontmatter@RRAPformat{\produce@RRAP{*#1\href{mailto:#2}{#2}}}\frontmatter@RRAPformat}
  {}{}
}%
\makeatother

\usepackage{ulem}
\usepackage{color}

\definecolor{violet}{rgb}{0.56, 0.0, 1.0}

\begin{document}

\preprint{AIP/123-QED}

\title{Multistability and Control in Ring Networks of Phase Oscillators with Frequency Heterogeneity and Phase Lag}
\author{Soomin Kim}
\altaffiliation{Electronic mail: thruminkim@gmail.com}%
\affiliation{Graduate School of Frontier Sciences,
The University of Tokyo, Chiba 277-8561, Japan
}%
\author{Hiroshi Kori}%
\altaffiliation{Electronic mail: kori@k.u-tokyo.ac.jp}
\affiliation{Graduate School of Frontier Sciences,
The University of Tokyo, Chiba 277-8561, Japan
}%

\date{\today}

\begin{abstract}
Many oscillator networks are multistable, meaning that different synchronization states are realized depending on the initial conditions. In this paper, we numerically analyze a ring network of phase oscillators, in which synchronous states with different wavenumbers are simultaneously stable. This model is an extension of the one studied in detail in previous studies by introducing inhomogeneities in the natural frequencies and the phase lag in the interaction, which are essential factors in the application. We investigate basin size distribution, which characterizes the size of the initial value set that converges to each synchronous state, showing that the basin size of synchronous states with higher wave-numbers broadens as the phase lag increases up to a certain extent. Weak inhomogeneities in the natural frequencies are also found to broaden the basin size of synchronous states with lower wave-numbers, i.e., more synchronous states. The latter result is seemingly counter-intuitive, but occurs because the higher wavenumber states are more vulnerable to inhomogeneity. Finally, we propose a control method that exploits inhomogeneity and phase lag to steer the system into a synchronized state with a specific wavenumber. This research furthers our understanding of the design principles and control of oscillator networks.

\end{abstract}

\maketitle

\section{Introduction}

Various natural and artificial systems are composed of a population of individual units forming a network.
Especially when those units are oscillators, it is known that those units interact with each other and show collective behavior called synchronization\cite{winfree67,kuramoto84,winfree01,pikovsky01}. Synchronization has been observed in a wide variety of system such as group of fireflies\cite{buck88}, chorus of frogs\cite{aihara2014spatio}, firing in neurons\cite{cossart2003attractor}, and power grids\cite{menck2014dead}. For comprehensive reviews of the Kuramoto framework and synchronization in oscillator networks, see Refs.~\cite{Acebron2005Kuramoto,Arenas08,DorflerBullo2014Survey}

Synchronization is essential for proper functioning of oscillator networks. The required synchronization patterns vary depending on the target and situation. For example, the heart functions as a pump by having numerous cardiac muscle cells beat in a nearly in-phase synchronization pattern.
However, the heart can enter a dynamic state with a complex spatial pattern known as ventricular fibrillation, which can lead to loss of function and death \cite{glass2005multistable}.
Thus, the heart is a multistable system with undesirable synchronization patterns. In contrast, the situation is different in locomotion. Many animals exhibit multiple gaits, generated by neural networks known as central pattern generators (CPGs) \cite{bucher2015central}. The synchronization patterns of CPGs are believed to switch flexibly through inputs from the central nervous system and feedback from external sensory inputs \cite{briggman2008multifunctional}. Therefore, CPGs can also be considered multistable systems in a broad sense.

The multistability of oscillator networks has been extensively studied.
A particularly transparent setting is a ring of coupled phase oscillators, which already supports multiple coexisting phase-locked patterns and allows systematic comparisons across attractors. \cite{canavier1999control,Ermentrout1985Rings} 
In such multistable systems, basin volume provides a practical, global measure of how likely each attractor is to be reached from random initial conditions. \cite{Menck2013BasinStability} 
Building on this viewpoint, Wiley and Strogatz numerically investigated large oscillator rings and reported that the basin-size distribution of uniformly twisted (winding-number) states is approximately Gaussian. \cite{wiley2006size} 
Subsequent work revisited this question from complementary angles: Delabays \textit{et al.} proposed an efficient numerical protocol to estimate basin volumes, \cite{delabays2017size} Zhang and Strogatz uncovered intricate basin geometry, \cite{zhang2021basins} and Groisman \textit{et al.} provided further numerical and analytical evidence resolving the Gaussian scaling and clarifying its underlying dynamical mechanism. \cite{Groisman2025SyncBasinResolved}
Despite the model’s simplicity, these studies highlight that the global multistable structure of oscillator rings can be remarkably rich, making them a useful testbed for understanding multistability and control in oscillator networks.

A central goal in the study of coupled-oscillator networks is not only to characterize which synchronized patterns exist, but also to understand their accessibility and to develop ways to select a desired one. 
In multistable systems, this amounts to asking how basin geometry and robustness can be shaped so that (i) an intended pattern becomes more likely to occur and (ii) trajectories can be reliably steered to a prescribed attractor. 
Because these questions are inherently global and nonlinear, even relatively simple models can reveal nontrivial design principles. 
In the following, we pursue these goals in a ring of phase oscillators and use the insights gained to construct a practical control strategy.

In this study, we investigate multistability and basin structure in a ring network of phase oscillators and develop a control strategy to select a desired synchronized pattern. 
Motivated by applications, we extend the classical ring model in two directions. 
First, we allow oscillators to have heterogeneous natural frequencies, an essential ingredient in many biological and chemical systems. 
Second, we adopt a Sakaguchi--Kuramoto-type coupling in which the interaction includes a constant phase lag \cite{sakaguchi1986soluble}. 
Such a phase lag is known to break gradient structure and to influence entrainment and wave formation in spatially extended oscillator media \cite{sakaguchi88b}. 
We then quantify how these two modeling ingredients reshape the accessibility of coexisting phase-locked states, as captured by basin-size statistics, and use the resulting robustness differences to propose a parameter-based control protocol.

The paper is organized as follows: In Sec.~II we introduce the model; in Sec.~III we quantify basin-size statistics and robustness and then propose a control protocol; and in the Appendix we summarize the linear stability analysis for the homogeneous twisted states.

\section{Model}
\begin{figure}
 \includegraphics[width=7.5cm]{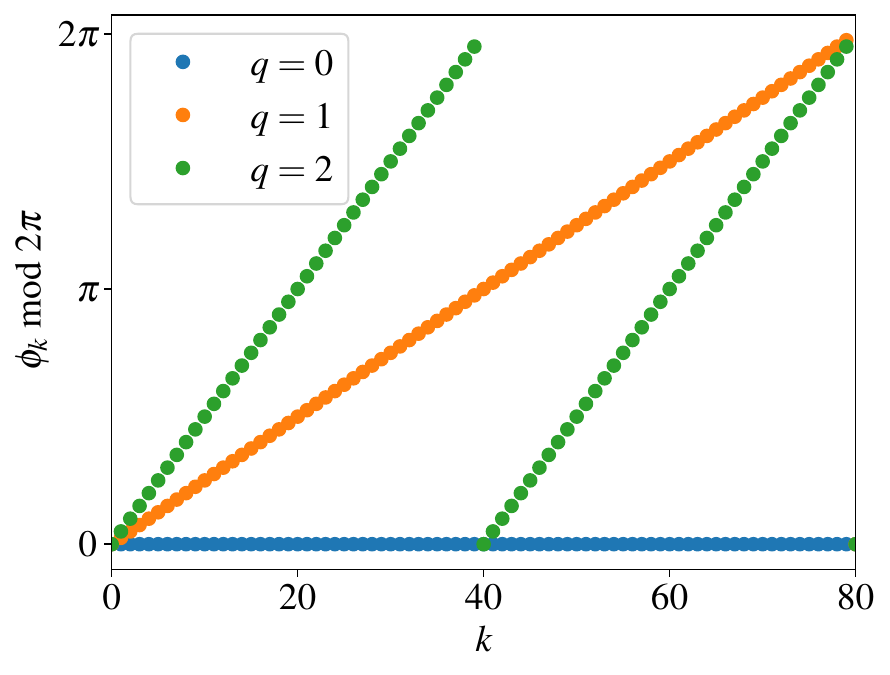}
 \caption{
Examples of uniformly $q$-twisted states defined by Eq.~\eqref{twist}. 
In a $q$-twisted state, the phase increment between neighboring oscillators is constant, 
$\phi_{k+1}-\phi_k = 2\pi q/N$, so that the phase winds by $2\pi q$ over one circuit of the ring.
}
\label{fig:twist}
\end{figure}

Our study is motivated by a predecessor work by Wiley and Strogatz \cite{wiley2006size}.
They investigated
a ring network of $N$ identical phase oscillators, each coupled with $l$ neighbors on both sides. Specifically, the dynamical equation is given as
\begin{equation}
\dot{\phi_{k}}=\omega+\sum_{j=k-l}^{k+l}\sin(\phi_{j}-\phi_{k}),
\label{model0}
\end{equation}
where $\omega$ and $\phi_k$ are the frequency and the phase of oscillator $k$ ($k=1,2,\ldots,N$), respectively, and $\phi_{k + m N}=\phi_k$ for $m=\pm 1, \pm 2, \ldots$. We may set $\omega=0$ without loss of generality because this treatment is equivalent to the variable change $\phi_k \to \phi_k-\omega t$ for $k=1,2,\ldots,N$.
By substituting Eq.~\eqref{twist} into Eq.~\eqref{model0}, one can confirm that this system has the following synchronized states:
\begin{equation}
 \phi_{k}=\Omega t+\frac{2 \pi q k}{N} + C,
\label{twist}
\end{equation}
where $\Omega$ is the synchronized frequency, $C$ is an arbitrary constant, and $q$ ($-\frac{N}{2}<q \le \frac{N}{2}$) is an integer called a winding number. When $k$ is regarded as a spatial axis, the wavenumber of the state is given as $2\pi q$. In this particular model, we have $\Omega=\omega$.
The state with $q$ is referred to as the uniformly $q$-twisted state. Some examples are shown in Fig.~\ref{fig:twist}. We also refer to the state with $q=0$ as the in-phase state.
Wiley and Strogatz defined the size of the basin of each $q$-twisted state as the probability that the system converges to the $q$-twisted state from fully random initial conditions.

In this study, as a variant of the original model \eqref{model0}, we consider
\begin{equation}
 \label{SKmodel}
  \dot{\phi_{k}}=\omega_{k}+ \sin(\phi_{k-1}-\phi_{k}+\alpha) + \sin(\phi_{k+1}-\phi_{k}+\alpha).
\end{equation}
where $\omega_k$ is the natural frequency of oscillator $i$ and $\alpha$ is the phase offset in the interaction function. For $\omega_k=\omega$ for $k=1,\ldots,N$, we have the homogeneous $q$-twisted states, given by Eq.~\eqref{twist}, where $\Omega=\omega+2 \sin \alpha \cos(\frac{2 \pi q}{N})$.
The phase oscillators with this particular interaction function was first consider by Sakaguchi and Kuramoto \cite{sakaguchi88b}.
 It is well known that $\alpha$ significantly affects synchronization properties and nonzero $\alpha$ is responsible for the generation of wave patterns in the presence of a pacemaker region in spatially extended systems \cite{kuramoto84,sakaguchi88b,blasius05}. Therefore, we expect that the basin size distribution depends critically on $\alpha$.
Natural frequency $\omega_k$ follows the normal distribution $\mathcal N(\omega, \sigma^2)$. The mean does not affect synchronization properties at all and is set to $\omega=0$ without loss of generality. In contrast, the effect of $\sigma$ is significant. For example, if $\sigma$ is sufficiently large, oscillators with natural frequencies that deviate considerably from the mean will be not synchronized. If the distribution is sufficiently narrow, all oscillators will be synchronized. Even in this case, the basin size distribution may depend on $\sigma$.
For simplicity, we focus on $k=1$ and do not investigate the effect of non-local coupling; i.e.,  $l \ge 2$ in Eq.~\eqref{model0} in this study.

\section{Numerical analysis}
We perform numerical simulations of Eq.~\eqref{SKmodel} with $N=80$ to characterize basin-size statistics, robustness against heterogeneity, and parameter-based control. We focus on the regime where the homogeneous in-phase state ($q=0$) is linearly stable ($\sigma=0$) in the homogeneous system ($\sigma=0$). This holds for $-\pi/2<\alpha<\pi/2$ (Appendix~\ref{sec:stability}). Within this range, uniformly twisted states are stable for $|q|<N/4$ in the homogeneous system. Moreover, because the model is invariant under $\phi_k\to-\phi_k$, $\omega_k\to-\omega_k$, and $\alpha\to-\alpha$, ensemble-averaged statistics are symmetric in the sign of $\alpha$; we therefore restrict to $0\le\alpha<\pi/2$.

\subsection{Basin sizes}
We first focus on the size of the basin of phase-locking states.
In each run of simulation, we randomly and independently set $\omega_k$ and $\phi_k(0)$ for $i=1,\ldots,N$, following the normal distribution $\mathcal N(0,\sigma^2)$ and the uniform distribution in the range $[0, 2\pi)$, respectively. We then run a numerical simulation for 
$0\le t \le 10^5$.
Convergence to a phase-locking state is monitored using Kuramoto order parameter $r(t) = \frac{1}{N}|\sum_{j=1}^N e^{i \phi_j}|$. 
We terminate a run when $|r(t+1)-r(t)|<10^{-7}$, indicating that the collective coherence has become stationary.
In this case, we calculate the winding number $q$ by
\begin{equation}
 q=\frac{1}{2\pi}\sum_{k=1}^{N}\psi_k,
\end{equation}
where $\psi_k=\phi_{k+1}-\phi_k$ for $k=1,\ldots, N-1$ and $\psi_N=\phi_{N}-\phi_1$, which are defined in the range $(-\pi,\pi]$. We repeat this for $M=10^4$ times to obtain the distribution of realized $q$.

\begin{figure}
 \includegraphics[width=7.5cm]{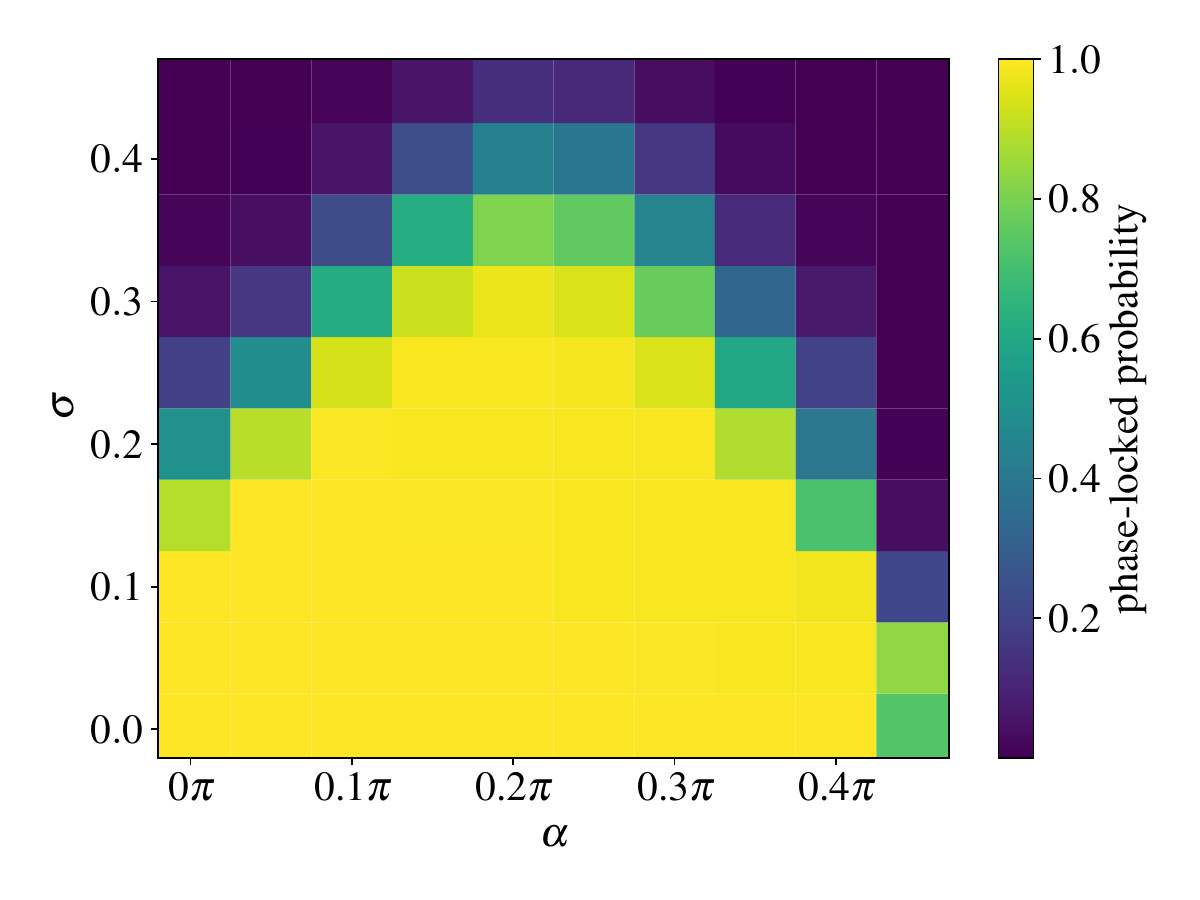}
 \caption{
Phase-locking probability on the $(\alpha,\sigma)$ parameter plane for Eq.~\eqref{SKmodel} ($N=80$). For each parameter pair, we perform $M=10^4$ independent runs with random initial phases and random frequency realizations; the color indicates the fraction of runs that converge to a phase-locked state (i.e., the number of phase-locked outcomes divided by $M$). 
A run is judged to be phase-locked when the Kuramoto order parameter becomes stationary, $|r(t+1)-r(t)|<10^{-7}$ (see Sec.~III\,A). 
The plot is shown as a rectangular tiling to emphasize the shape of the phase-locking region.
}
\label{fig:map}
\end{figure}

In Fig. \ref{fig:map}(a), we show the fraction of runs that converge to a phase-locked state (i.e., the number of phase-locked outcomes divided by $M$) for given parameter values.
We observe that for small $\sigma$, the system converges to a phase-locking state for all or most runs. As $\sigma$ increases, the convergence occurs less likely, as naturally expected. We also observe that convergence is facilitated and impeded as $\alpha$ increases for $0 \leq \alpha \lesssim 0.25 \pi$ and $0.25 \pi \lesssim \alpha \leq 0.45 \pi$, respectively. The system synchronizes most robustly for $\alpha \simeq 0.20 \pi$.

\begin{figure}
 \includegraphics[width=7.5cm]{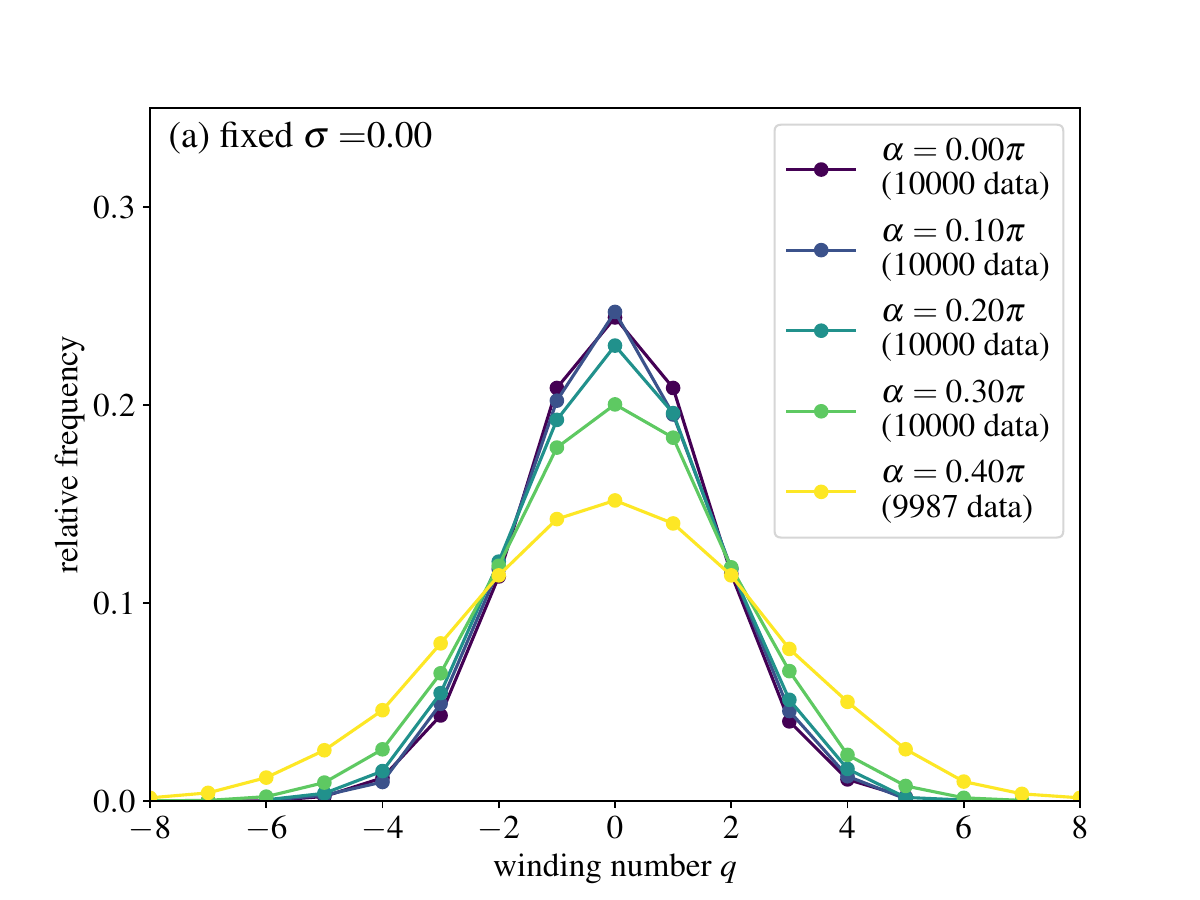}
 \includegraphics[width=7.5cm]{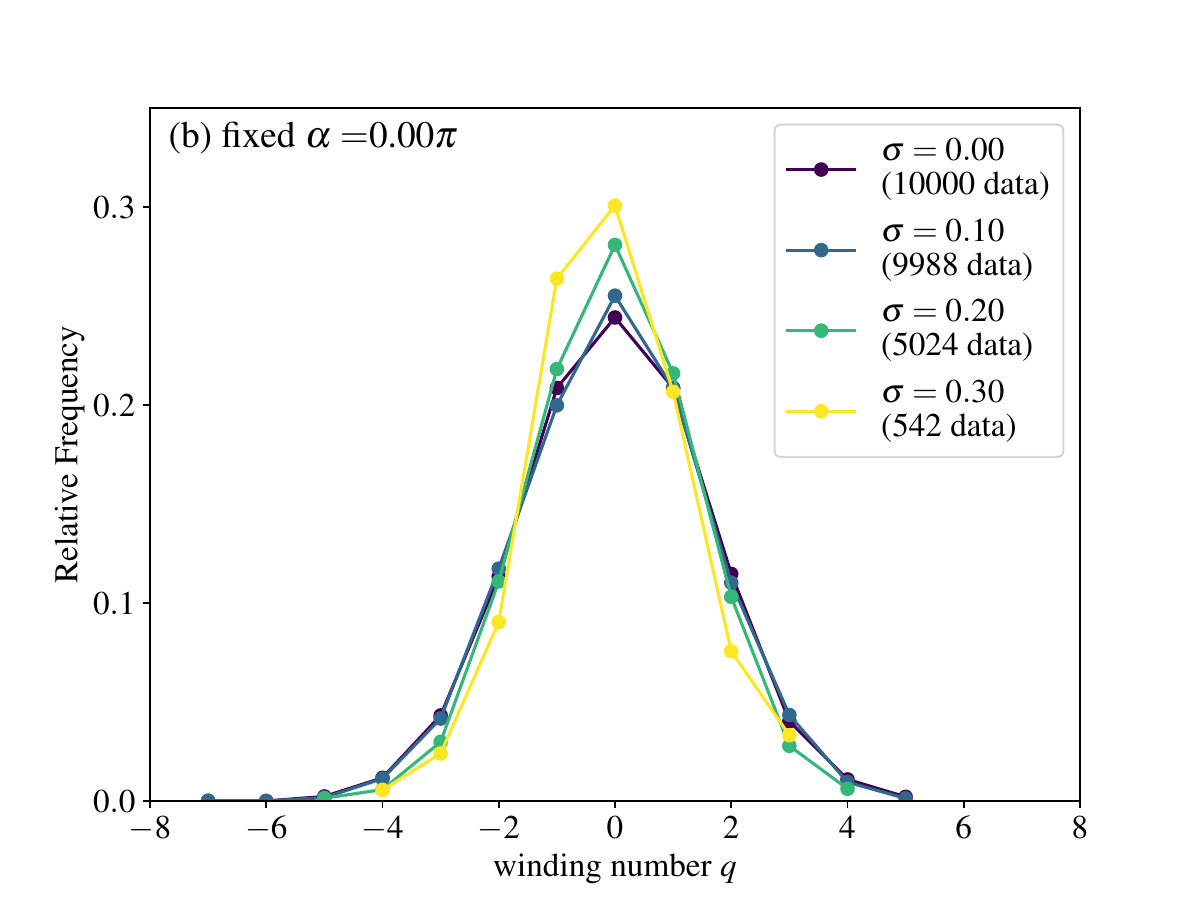}
 \includegraphics[width=7.5cm]{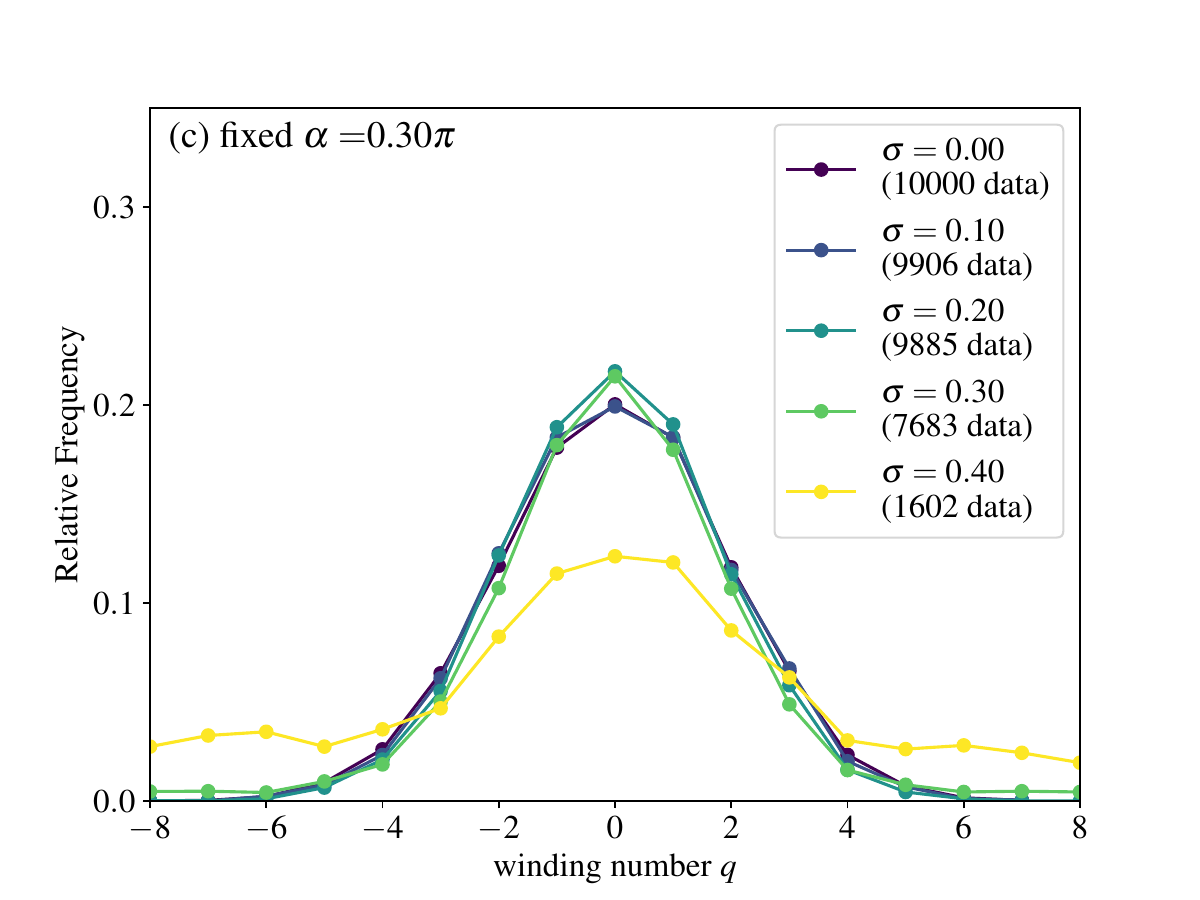}
 \includegraphics[width=7cm]{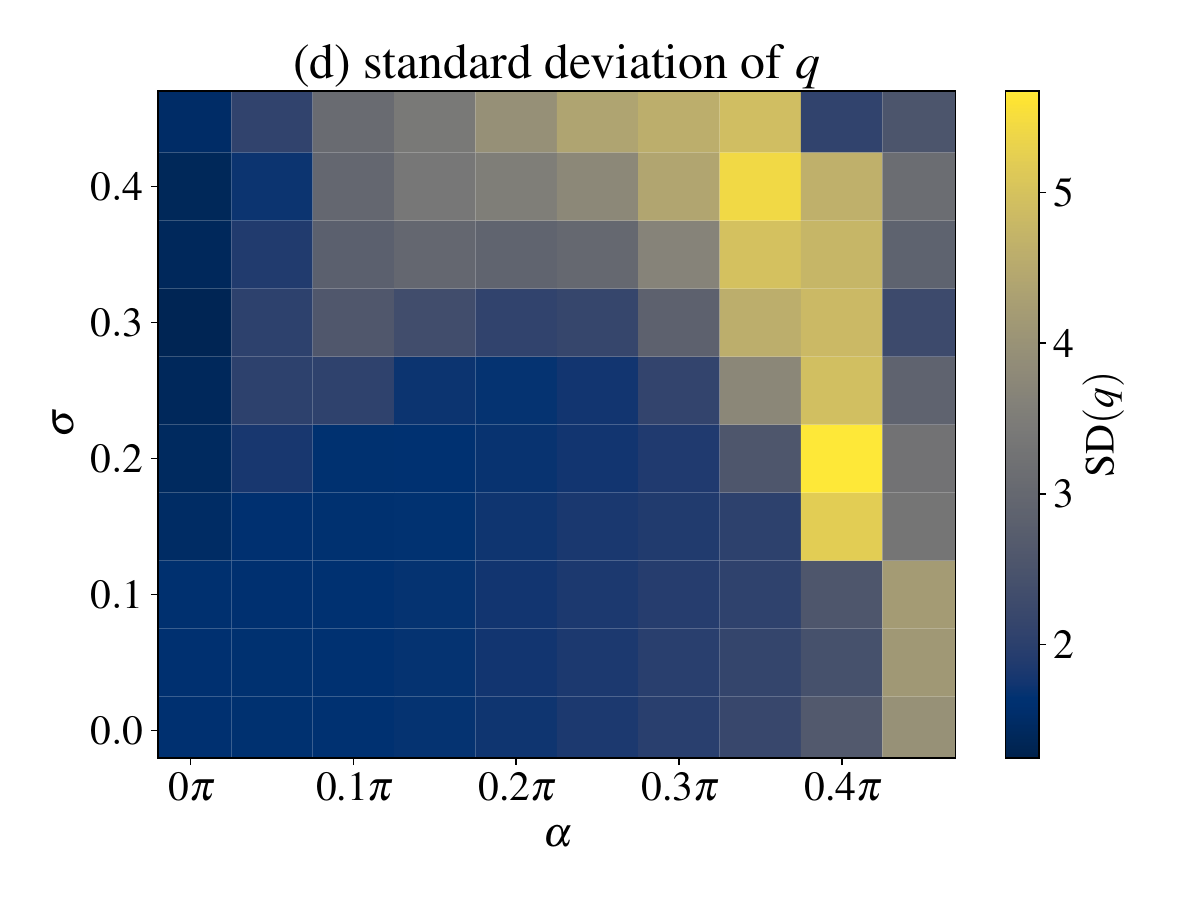}
\caption{
Basin-size statistics of phase-locked $q$-twisted states, obtained from random initial phases and random frequency realizations (Sec.~III\,A, $N=80$).
(a) Empirical distribution of the winding number $q$ for $\sigma=0$ and several values of $\alpha$.
(b) Distribution of $q$ for $\alpha=0$ and several values of $\sigma$.
(c) Distribution of $q$ for $\alpha=0.3\pi$ and several values of $\sigma$.
In (a)--(c), frequencies are shown as relative frequencies \emph{conditioned on phase-locking outcomes}.
(d) Standard deviation of the $q$ distribution as a function of $(\alpha,\sigma)$, summarizing the broadening/narrowing trends in (a)--(c).
}
\label{fig:histo}
\end{figure}

We next focus on the distribution of the winding number $q$ of obtained phase-locking states. 
Figure \ref{fig:histo}(a), which is for different $\alpha$ values with fixed $\sigma=0$, shows that the distribution becomes wider as $\alpha$ increases while keeping its unimodality.
This property holds approximately true for $\sigma>0$ (result not shown).
This indicates that convergence to a twisted state with larger $|q|$ values is facilitated as $\alpha$ increases. On the other hand, the $\sigma$ dependency is rather complicated.
For $\alpha=0$, as shown in \ref{fig:histo}(b), the distribution becomes narrower as $\sigma$ increases. This suggests that the frequency heterogeneity breaks the states with large $|q|$ values. 
For $\alpha=0.3\pi$, as shown in Fig. \ref{fig:histo}(c), a similar tendency is observed for $\sigma \le 0.2$, but for $\sigma > 0.2$, conversely, the distribution becomes wider as $\sigma$ increases. 
In any cases, the distributions look like Gaussian; this property seems robust against the frequency heterogeneity and the phase lag in the interaction function.
As a summary, we show the standard deviation of the $q$ distribution for each parameter set $(\alpha,\sigma)$ in Fig. \ref{fig:map}(b).
We can confirm that the standard deviation increases as $\alpha$ increases for any $\sigma$ values, whereas the dependence on $\sigma$ is not necessarily monotonic.

\subsection{Robustness of twisted states against the frequency heterogeneity}
\begin{figure}[t]
 \includegraphics[width=6.5cm]{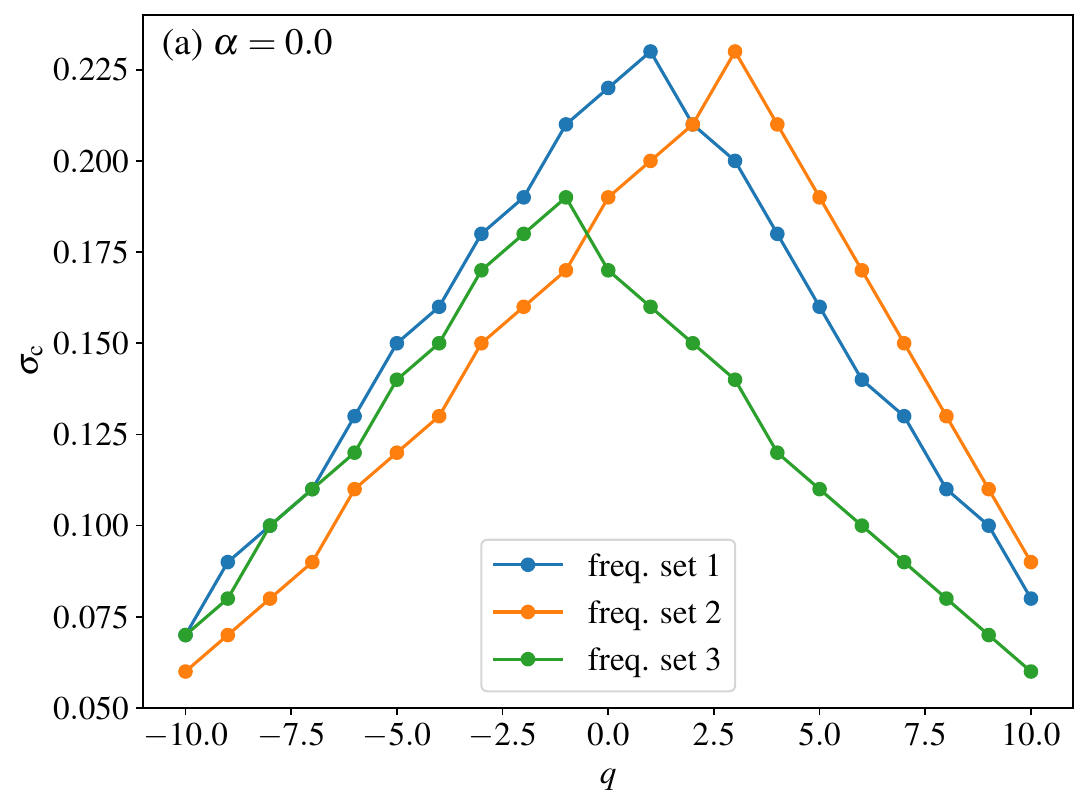}
 \includegraphics[width=6.5cm]{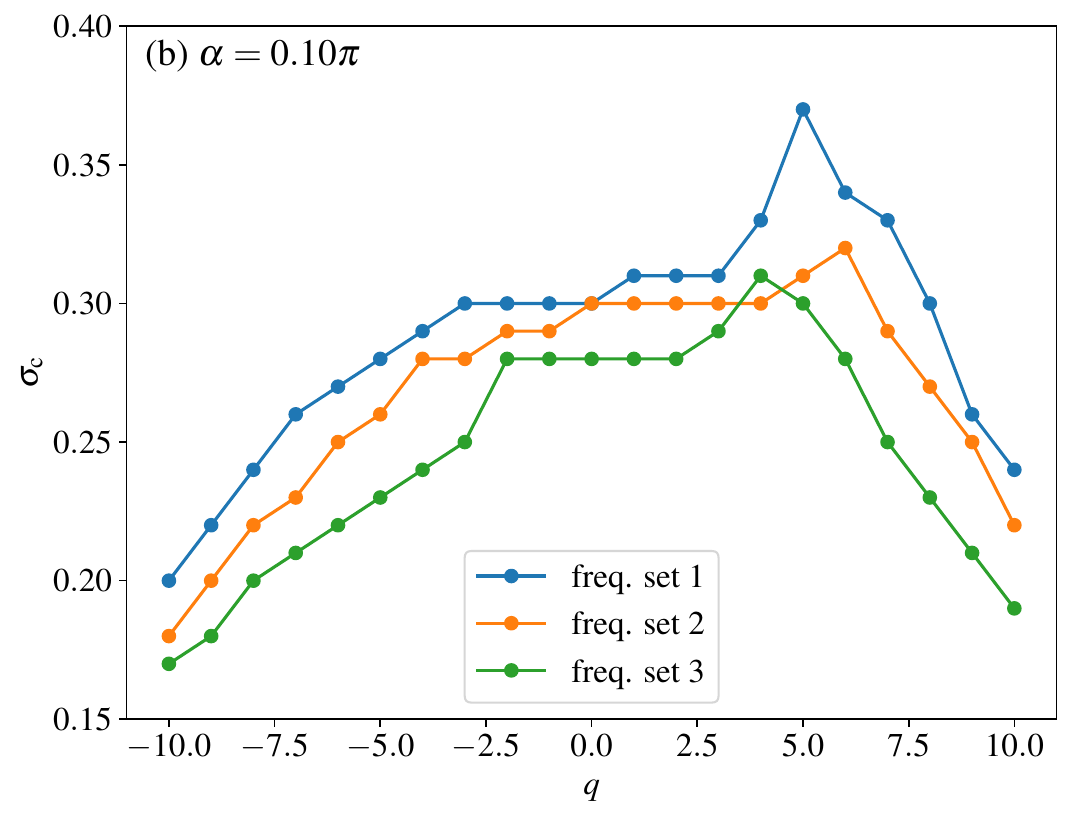}
 \caption{
Critical heterogeneity $\sigma_{\rm c}(q)$ for each $q$-twisted state for three representative frequency sets (Sec.~III\,B, $N=80$).
For each $q$, the simulation starts from the homogeneous $q$-twisted state and $\sigma$ is increased in steps of $\Delta\sigma=0.01$ until the trajectory no longer converges to a phase-locked state with the same winding number $q$.
(a) $\alpha=0$. (b) $\alpha=0.1\pi$.
}
 \label{fig:sigma_crit_each}
\end{figure}

\begin{figure}[t]
 \includegraphics[width=8cm]{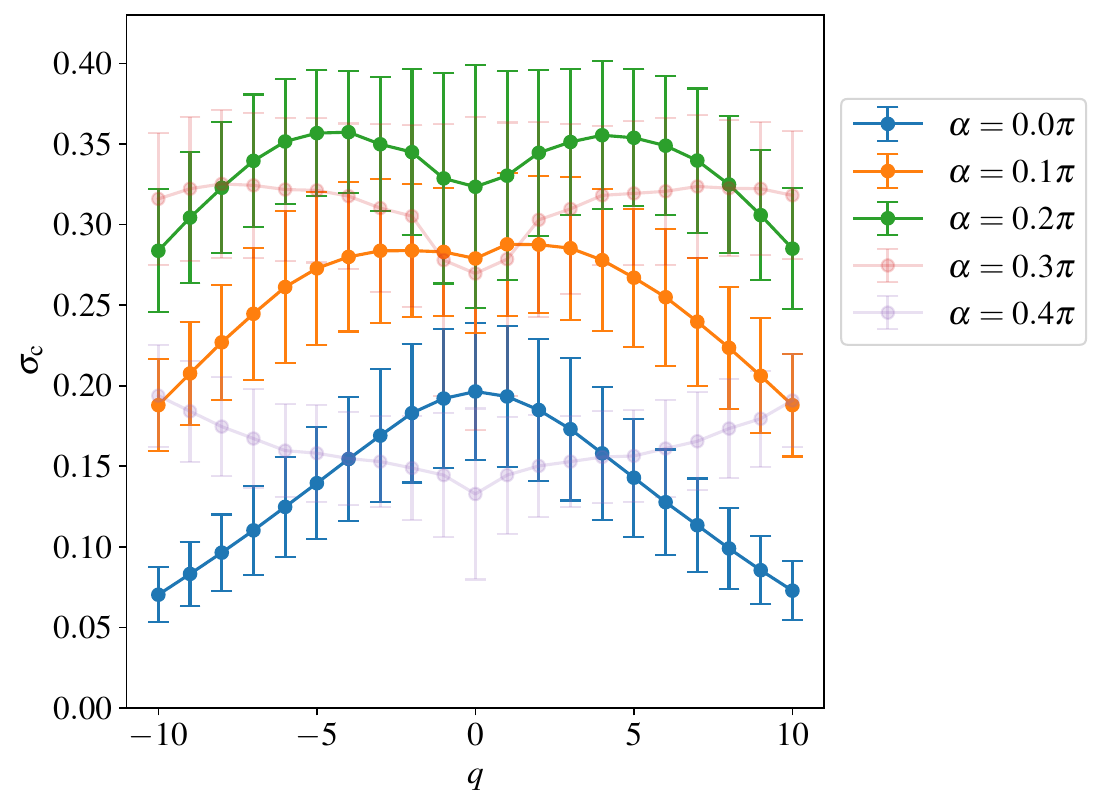}
 \caption{
Ensemble statistics of the critical heterogeneity $\sigma_{\rm c}(q)$ over 100 independent frequency realizations $\{\hat\omega_k\}$ (Sec.~III\,B).
Symbols show the mean of $\sigma_{\rm c}(q)$ and error bars indicate the standard deviation across realizations, for each value of $\alpha$.
}
 \label{fig:sigma_crit}
\end{figure}

To understand the effect of the frequency heterogeneity on the multi-stability, we measure the critical $\sigma$ values, denoted as $\sigma_{\rm c}(q)$, above which the stable $q$-twisted state no longer exist. The procedure to obtain $\sigma_{\rm c}(q_1)$ is as follows.
We first generate a frequency set, $\{\hat \omega_k\}$  $(i=1,\ldots,N)$, from the normal distribution $\mathcal N(0,1)$ and set $\{\omega_k\} = \{\sigma \hat \omega_k \}$. The initial condition is given by the uniformly twisted state, Eq.~\eqref{twist}, for $q=q_1$.
We then run a simulation for $\sigma=\Delta \sigma$, where $\Delta \sigma=0.01$. If the system converges to a phase-locked state and the obtained $q$ value is the same as $q_1$, we increase $\sigma$ by $\Delta \sigma=0.01$ and run a simulation without resetting the $\theta_k$ values. Otherwise, the current $\sigma$ value gives $\sigma_{\rm c}(q_1)$.
In Fig.~\ref{fig:sigma_crit_each}, we plot $\sigma_{\rm c}(q)$ for three different frequency sets with (a) $\alpha=0$ and (b) $0.1\pi$. We observe that $\sigma_{\rm c}(q)$ is unimodal and peaked at different $q$ values.
Comparing $\alpha=0$ and $0.1\pi$ cases, the latter tends to have a maximum $\sigma_{\rm c}(q)$ at a larger $q$ value. In Fig.~\ref{fig:sigma_crit}, 
we plot the mean and the standard deviation of $\sigma_{\rm c}(q)$ obtained from 100 different sets of $\{\hat \omega_k\}$. We observe that $\sigma_{\rm c}$ increases and decreases as $\alpha$ increases for $\alpha=0,0.1\pi, 0.2\pi$ and for $\alpha= 0.3\pi, 0.4\pi$, respectively. Thus, the phase lag up to some extent makes the synchronization robust against frequency heterogeneity.
For $\alpha=0$, the maximum $\sigma_{\rm c}(q)$ is located at $q=0$, which is the nearly in-phase state.
This indicates that, as heterogeneity is increased, states with larger values of $q$ tend to disappear earlier, and, statistically, the nearly in-phase state is most robust against the frequency heterogeneity. 
This result is consistent with the result shown in Fig. \ref{fig:histo}(a), in which large heterogeneity promotes the convergence to the nearly in-phase state.
For $\alpha>0$, the maximum $\sigma_{\rm c}(q)$ is located at $|q|>0$, thus the frequency heterogeneity promotes the convergence to twisted states with larger $|q|$ values. 

\subsection{Control}
\begin{figure}
 \includegraphics[width=8cm]{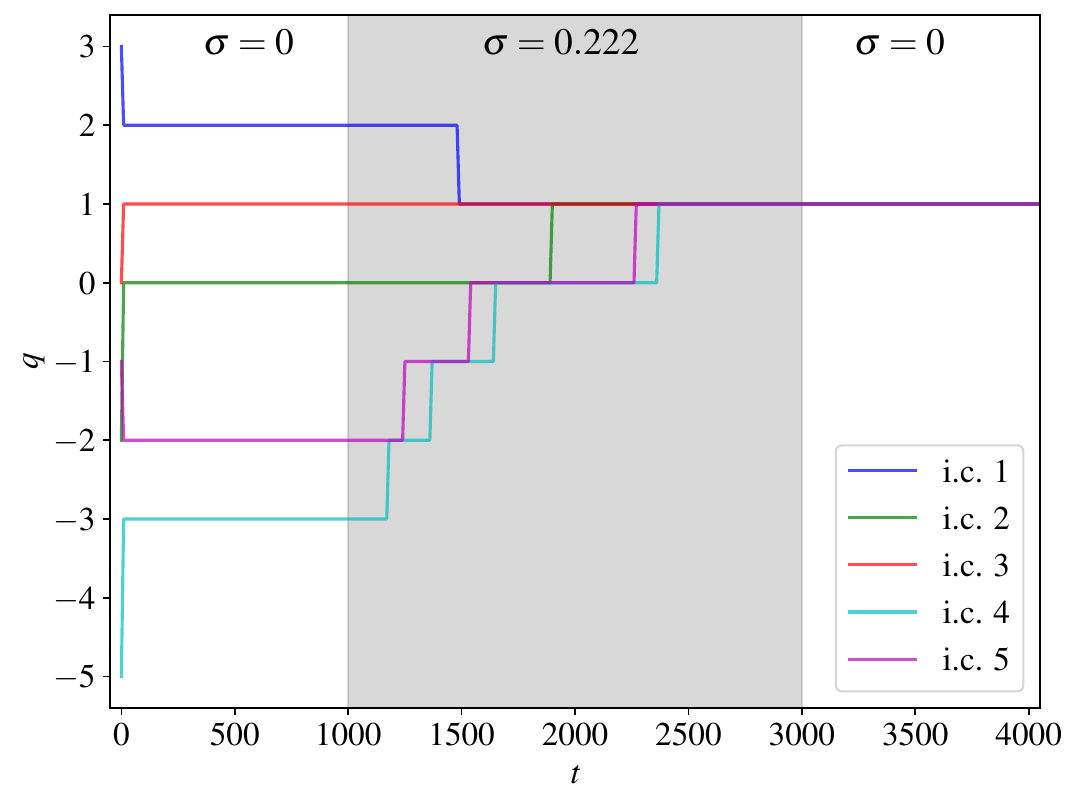}
 \caption{
Demonstration of parameter-switching control (Sec.~III\,C).
Time series of the winding number $q(t)$ for five random initial conditions ($N=80$), using frequency set~1 at $\alpha=0$.
The heterogeneity is switched from $\sigma=0$ to $\sigma=0.222$ during $1000\le t\le 3000$, which destabilizes all competing phase-locked states except the target $q=1$ state; after switching back to $\sigma=0$, the target twisted state persists.
}
 \label{fig:control}
\end{figure}

Our robustness results suggest a control strategy that does not require direct manipulation of individual phases. For a given realization of the frequency set $\{\hat\omega_k\}$, Fig.~\ref{fig:sigma_crit_each} shows that there can exist a narrow interval of $\sigma$ (for a chosen $\alpha$) where only a single winding number $q^*$ remains phase-locked while all other phase-locked twisted states have lost. This motivates the following parameter-switching protocol: (i) let the system evolve at baseline parameters (e.g., $\sigma=0$) until it approaches some phase-locked state; (ii) temporarily switch to parameters $(\alpha,\sigma)$ chosen so that only the target $q^*$ remains stable; (iii) after convergence to $q^*$, restore the baseline parameters, under which the target state remains stable. This protocol effectively removes undesired attractors by exploiting their different heterogeneity tolerances. As an example, we use frequency set 1 with $\alpha=0$, for which the $q=1$ twisted state becomes effectively monostable for $0.220\lesssim\sigma\lesssim0.225$ (Fig.~\ref{fig:sigma_crit_each}(a)). Figure~\ref{fig:control} shows time series of $q$ for five random initial conditions, where we set $\sigma=0.222$ for $1000\le t\le3000$ and $\sigma=0$ otherwise. Different $q$ values are obtained for $0\le t<1000$, but once heterogeneity is turned on, all trajectories are driven to the target $q=1$ state; importantly, the target persists after $\sigma$ is set back to zero.

\section{Summary and discussion}

We studied multistability in a nearest-neighbor ring of phase oscillators with both natural-frequency heterogeneity ($\sigma$) and phase lag ($\alpha$).
By sampling random initial phases and frequency realizations, we quantified basin sizes of phase-locked $q$-twisted states and found that increasing $\alpha$ generally broadens the winding-number distribution, promoting higher-wavenumber twisted states up to an intermediate range of $\alpha$.
The effect of heterogeneity is subtler: for small $\alpha$, increasing $\sigma$ preferentially destabilizes higher-wavenumber states, effectively enlarging the basins of low-wavenumber (more synchronous) states; for larger $\alpha$, the most robust winding number shifts to $q>0$.
To make this robustness explicit, we computed the critical heterogeneity $\sigma_{\rm c}(q)$ beyond which each $q$-twisted state loses phase locking. Across frequency realizations, $\sigma_{\rm c}(q)$ is typically unimodal and its maximizer shifts to larger $|q|$ as $\alpha$ increases, explaining the observed changes in basin-size statistics.
Finally, we leveraged these differences to propose a parameter-switching control protocol that transiently removes undesired attractors, thereby steering the system to a prescribed winding number without phase-level actuation.
Understanding basin sizes remains challenging because basin boundaries are shaped by the nonlinearility of the system. Even for $\alpha=\sigma=0$, basin statistics have been studied mainly numerically \cite{wiley2006size,delabays2017size,zhang2021basins}.
Our results indicate that both phase lag and heterogeneity can be useful design knobs for controlling multistable oscillator networks, and motivate further theoretical work to connect basin-size changes to underlying invariant-set geometry.

\section*{Acknowledgments}
This study was supported by the JSPS KAKENHI (Nos. JP21K12056, JP22K18384, and JP23H02796) to H.K.

\clearpage
\onecolumngrid
\appendix

\section{stability analysis of uniformly twisted states} \label{sec:stability}
Here, we conduct the stability analysis of the uniformly twisted states, given by Eq.~\eqref{twist}, for Eq.~\eqref{SKmodel} with $\omega_k=0$ (i.e., without frequency heterogeneity).
We describe the state as
\begin{equation}
 \phi_k = \Omega t + \frac{2\pi q k}{N} + \xi_k,
\end{equation}
where $\Omega=2 \sin \alpha \cos \frac{2\pi q}{N}$ and
$\xi_k$ is a small deviation from the the uniformly twisted state. By substituting this into Eq.~\eqref{SKmodel}, we obtain
\begin{align}
 \dot \xi_k &= \sin\left( \frac{2\pi q}{N} + \xi_{k-1} - \xi_{k} + \alpha \right)
  + \sin\left( - \frac{2\pi q}{N} + \xi_{k+1} - \xi_{k}+ \alpha \right)
 - 2 \sin \alpha \cos \frac{2\pi q}{N}\\
  &=  \sin\left(\frac{2\pi q}{N} +\alpha \right)\cos(\xi_{k-1} - \xi_{k})
 + \cos\left(\frac{2\pi q}{N} +\alpha \right)\sin(\xi_{k-1} - \xi_{k})\\
 & + \sin\left(-\frac{2\pi q}{N} +\alpha \right)\cos(\xi_{k+1} - \xi_{k})
 + \cos\left(-\frac{2\pi q}{N} +\alpha \right)\sin(\xi_{k+1} - \xi_{k})- 2 \sin \alpha \cos \frac{2\pi q}{N}.
\end{align}
By linearizing the equation for small $\xi_k$, we obtain
\begin{align}
 \dot \xi_k
  &=  \cos\left(\frac{2\pi q}{N} +\alpha \right)(\xi_{k-1} - \xi_{k})
 + \cos\left(-\frac{2\pi q}{N} +\alpha \right)(\xi_{k+1} - \xi_{k}).
\end{align}

The stability of this linear equation can be found by substituting the following expression into Eq.~\eqref{SKmodel}:
\begin{equation}
 \xi_k = e^{\lambda_p t + \frac{2 \pi p}{N} k},
\end{equation}
where $p$ ($p=0,1, \ldots, N-1$) is the wave number of the perturbation
and $\lambda_p$ is the eigenvalues of the stability matrix for the $q$-twisted state.
As a result, we obtain
\begin{align}
 \lambda_p =  -2 \left\{ \cos \alpha \cos \frac{2\pi q}{N} \left(1- \cos \frac{2 \pi p}{N} \right) +
i \sin \alpha \sin \frac{2\pi q}{N} \sin \frac{2 \pi p}{N} \right\}
\label{eigenvalues}
\end{align}
Trivially, $\lambda_0=0$, which exists because of the translational symmetry of the system.
If the real parts of the remainder eigenvalues are all negative, then the $q$-twisted state is stable. This is the case for $-\frac{\pi}{2} < \alpha < \frac{\pi}{2}$ and $|q| < \frac{N}{4}$.
For $\frac{\pi}{2} < |\alpha| < \pi$, the $q$-twisted state with $\frac{N}{4} < |q| < \frac{N}{2}$ are stable.
From Eq.~\eqref{eigenvalues}, it is clear that the real part is proportional to $\cos \alpha$, thus for any $\alpha$, the ratio between the real parts of the eigenvalues of two different $q$-twisted states are the same.

\end{document}